\documentclass[aps,prl,twocolumn,showpacs]{revtex4}
\usepackage{amssymb}

\usepackage{graphicx}

\begin{document}


\title{Low Temperature Specific Heat of Optimally Doped BaFe$_{2-x}$$TM_x$As$_{2}$ ($TM$ = Co and
Ni) Single Crystals: Constraint on the Pairing Gap}

\author{Bin Zeng, Gang Mu, Bing Shen, Peng Cheng, Huiqian Luo, Huan Yang, Lei Shan, Cong Ren and Hai-Hu
Wen$^1$$^{\star}$}

\affiliation{$^1$National Laboratory for Superconductivity,
Institute of Physics and Beijing National Laboratory for Condensed
Matter Physics, Chinese Academy of Sciences, P.O. Box 603, Beijing
100190, China}

\begin{abstract}
Low temperature specific heat has been measured in optimally doped
and highly overdoped non-superconducting BaFe$_{2-x}$TM$_x$As$_{2}$
($TM$ = Co and Ni) single crystals. By using the data of the
overdoped samples, we successfully removed the phonon contribution
of the optimally doped ones, and derived the electronic specific
heat coefficient $\gamma_e$. Remarkably, we found a continuing
temperature dependent $\gamma_e(T)$ which follows the quadratic
relation $\gamma_e=\gamma_0+\alpha T^2$ in the low temperature
limit. Together with the very small residual term $\gamma_0$, linear
magnetic field dependence of $\gamma_e$, it is concluded that there
are either small segments of nodal lines, or point-like nodes in
these samples.
\end{abstract}

 \maketitle

The discovery of superconductivity above 50 K in iron pnictides has
added a new member in the family of unconventional high temperature
superconductors.\cite{Kamihara2008} One of the key issues here is
about the superconducting pairing mechanism. Theoretically it was
suggested that the pairing may be established via inter-pocket
scattering of electrons between the hole pockets (around $\Gamma$
point) and electron pockets (around M point), leading to the
so-called S$^\pm$ pairing manner.\cite{Mazin,Kuroki,LeeDH,LiJX}
Experimental results about the pairing symmetry remain highly
controversial leaving the perspectives ranging from S$^{++}$-wave,
to S$^\pm$ and to
d-wave.\cite{LaFePOCarrington,LaFePO,HDing,Hashimoto,Grafe,LuoXG,Prozorov,Matsuda,Sato,MuG}
Although evidence for a nodal gap has been accumulated in
LaFePO\cite{LaFePOCarrington,LaFePO} and Ba(FeAs$_{1-x}$P$_x$)$_2$
systems\cite{Matsuda}, in the charge doped 122, the experimental
data point to the existence of isotropic gaps, especially in the
optimally doped samples.\cite{HDing,LuoXG,Prozorov} The penetration
depth measurements indicate a quadratic temperature dependence
$\Delta \lambda \propto 1-(T/T_c)^2$ when the measuring inductive
current is flowing along FeAs-plane.\cite{Prozorov} And it becomes
more linear like when the detecting current has a c-axis component,
indicating the possibility of nodes along c-axis but nodeless along
FeAs-plane.\cite{Prozorov2} This is further strengthened by the
recent thermal conductivity measurements which suggest the possible
existence of a horizontal nodal line.\cite{Taillefer} In this
Letter, we present the data of low temperature specific heat (SH) in
BaFe$_{2-x}$TM$_x$As$_{2}$ ($TM$ = Co and Ni) single crystals. A
unique quadratic temperature dependence $\gamma_e = \gamma_0 +
\alpha T^2$ was found in the low-T limit. This is different from the
expectation for a d-wave superconductor in the clean limit
$C_e\propto T^2$. Combining with the very small $\gamma_0$ and the
linear field dependence of $\gamma_e$, it is tempting to conclude
the existence of small segments of nodal lines or point-like nodes,
being consistent with the recent five band tight binding
calculations.\cite{Graser}

The Co- and Ni-doped BaFe$_2$As$_2$ single crystals were grown by
the self-flux method\cite{FangLei,LuoHQ}. In order to get the phonon
contribution of SH, we also measured an overdoped
non-superconducting sample as the reference for each kind of dopant.
The typical dimension of the single crystals for specific heat
measurements was 3$\times$ 3$\times$ 0.4 mm$^3$. The SH measurements
were done with the thermal relaxation method on the Quantum Design
instrument physical property measurement system (PPMS) with the
temperature down to 1.8 K and magnetic field up to 9 T. To improve
the resolution, the field dependence of the sensor on the measuring
chip was calibrated in advance.

\begin{figure}
\includegraphics[width=8cm]{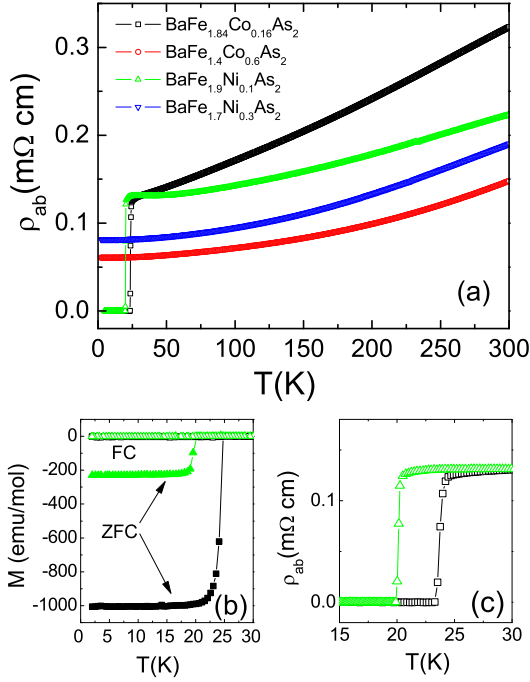}
\caption {(color online) (a) Temperature dependence of resistivity
of the optimally doped and highly overdoped non-superconducting
BaFe$_{2-x}TM_x$As$_2$ ($TM$ = Co and Ni) single crystals at zero
field. (b) Temperature dependence of dc magnetization for the two
superconducting samples. The enlarged view of resistivity for the
two superconducting samples is shown in (c).} \label{fig1}
\end{figure}

In Fig.1(a), we show the temperature dependence of resistivity for
four samples: BaFe$_{2-x}$Co$_{x}$As$_2$ with $x=0.16$ and 0.60, and
BaFe$_{2-x}$Ni$_{x}$As$_2$ with $x=0.10$ and 0.30, respectively. The
onset transition temperatures determined at $\rho = 95\%\rho_n$ for
BaFe$_{1.84}$Co$_{0.16}$As$_2$ and BaFe$_{1.9}$Ni$_{0.1}$As$_2$ are
24.2 K and 20.5 K, respectively, which are close to the optimal
doping points. The sharp transitions near $T_c$ in the dc
magnetization and resistivity measurements indicate the good quality
of the samples. No superconductivity was detected in
BaFe$_{1.4}$Co$_{0.6}$As$_2$ and BaFe$_{1.7}$Ni$_{0.3}$As$_2$ down
to 1.8 K.

\begin{figure}
\includegraphics[width=8cm]{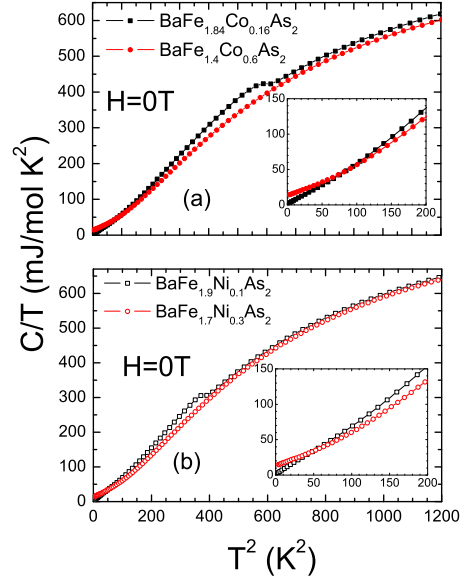}
\caption {(color online) Raw data of the temperature dependence of
specific heat for the BaFe$_{2-x}$TM$_x$As$_{2}$ ($TM$ = Co and Ni)
single crystals. The enlarged views of the same data are presented
as insets in (a) and (b), respectively.} \label{fig2}
\end{figure}

In Fig.2, we present the SH of the superconducting and
non-superconducting samples under zero field for each dopant. Sharp
and clear SH anomalies near $T_c$ for the samples
BaFe$_{1.84}$Co$_{0.16}$As$_2$ and BaFe$_{1.9}$Ni$_{0.1}$As$_2$ can
be seen. For all samples investigated here, the Schottky anomalies
are negligible. For each dopant, one can see that the SH in the
normal state for the superconducting and non-superconducting samples
share a very similar temperature dependence. This allows us to use
the phonon part of SH of the non-superconducting sample as the
reference, and subtract it safely for the superconducting one.

In order to extract the electronic SH of the superconducting
samples, people usually apply a magnetic field to destroy the
superconductivity, and measure the SH of the normal state. For the
iron-based superconductors, however, the upper critical field is too
high to be reached.\cite{Stewart} So we employ the method suggested
in the literatures\cite{Keimer,Ronning}. Here we take the overdoped
non-superconducting sample as the references. Suppose that the
phonon contributions of the non-superconducting sample is
$C_{ph}^{N}(T)$, which can be obtained by subtracting the linear
electronic term from the total SH, and that of the superconducting
sample is $C_{ph}^{S}(T)$, we naturally have $C_{ph}^{S}(T)=a\cdot
C_{ph}^{N}(b\cdot T)$. Here $a$ and $b$ are fitting parameters which
should be close to unity. Using a least-squares fit of our data
above T$_c$, we determined $a$ and $b$ to be 1.000, 1.013 for the
Co-doped sample, and 0.98, 0.99 for the Ni-doped one, indicating
that the phonon contributions of the superconducting sample and the
overdoped non-superconducting one are indeed very close to each
other. Although the antiferromagnetic (AF) spin fluctuation is very
strong in the optimally doped sample and negligible in the overdoped
one\cite{Imai}, the SH concerns mainly the Q = 0 scattering and is
sensitive only to the quasiparticle density of states (DOS) at
$E_F$. The entropy conservation (not shown here) obtained by this
treatment supports above argument. The electronic SH of the
superconducting sample can then be obtained through

\begin{equation}
C_{e}^{S}(T)=C_{tot}^{S}(T)-C_{ph}^{S}(T) =C_{tot}^{S}(T)-a\cdot
C_{ph}^{N}(b\cdot T),\label{eq:3}
\end{equation}
where $C_{tot}^{S}(T)$, $C_{e}^{S}(T)$ are the total and the
electronic SH of the superconducting sample, respectively.

\begin{figure}
\includegraphics[width=8cm]{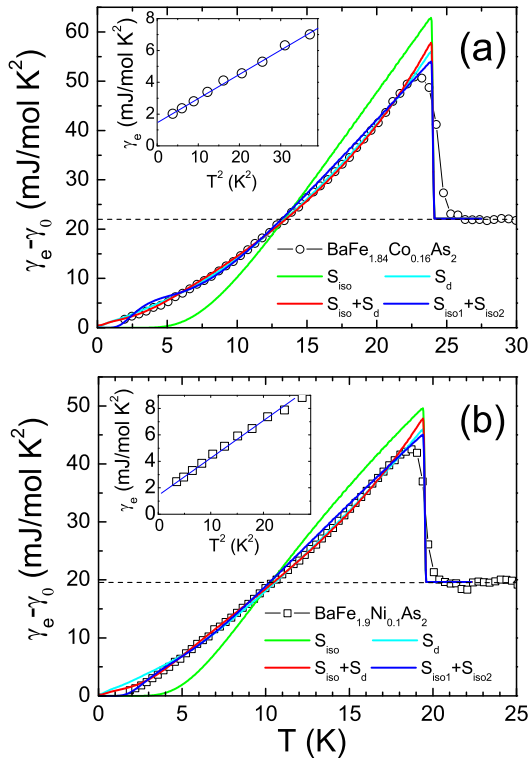}
\caption {(color online) The electronic specific heat of the
optimally doped superconducting samples (a)
BaFe$_{1.84}$Co$_{0.16}$As$_{2}$ and (b)
BaFe$_{1.9}$Ni$_{0.1}$As$_2$. These data were obtained by removing
the phonon contribution (see text). The insets show the enlarged
view of the low temperature specific heat, a quadratic temperature
dependence $\gamma_e=\gamma_0+\alpha T^2$ can be clearly seen. Four
different models: single isotropic s-wave gap ($S_{iso}$), single
anisotropic gap with a d-wave feature ($S_d$), mixture of two:
$S_{iso}+S_d$ and $S_{iso1}+S_{iso2}$ were used to fit the data. }
\label{fig3}
\end{figure}

The obtained electronic SH for Co- and Ni-doped samples are shown in
Fig.3. Surprisingly, we found a continuing temperature dependent
electronic SH coefficient $\gamma_e(T)$, which follows the quadratic
relation $\gamma_e=\gamma_0+\alpha T^2$ in the low-T limit (see the
insets of Fig.3). The parameters of $\gamma_0$ were determined from
the fitting: 1.53 mJ/mol K$^2$ and 1.49 mJ/mol K$^2$, for the
superconducting Co- and Ni-doped samples, respectively, which are
rather small compared to the values reported in literatures,
indicating the cleanness of our samples. In order to get a
comprehensive understanding, we used the BCS formula to fit our data

\begin{eqnarray}
\gamma_\mathrm{e}'=\frac{4N_F}{k_BT^{3}}\int_{0}^{+\infty}\int_0^{2\pi}\frac{e^{\zeta/k_BT}}{(1+e^{\zeta/k_BT})^{2}}(\varepsilon^{2}+\nonumber\\
\Delta^{2}(\theta,T)-\frac{T}{2}\frac{d\Delta^{2}(\theta,T)}{dT})\,d\theta\,d\varepsilon,
\end{eqnarray}

where $\gamma'_e=\gamma_e-\gamma_0$,
$\zeta=\sqrt{\varepsilon^2+\Delta^2(T,\theta)}$, N$_F$ the DOS at
the Fermi energy. We use four different gap structures to fit the
data: single isotropic s-wave gap ($S_{iso}$), single anisotropic
gap with a d-wave feature ($S_d$: $\Delta = \Delta_0 | cos2 \theta
|$), mixture of two: $S_{iso}+S_d$ and two isotropic gaps
$S_{iso1}+S_{iso2}$. In the latter two cases, the $\gamma_e$ was
calculated through a linear combination of the two components. The
fitting curves are shown by the colored lines in Fig.3, and the
fitting parameters are listed in Table I and Table II, respectively.
One can see that the single isotropic s-wave model can not fit the
data at all, while the fittings with $S_d$, $S_{iso}+S_d$ and
$S_{iso1}+S_{iso2}$ can roughly describe the data. However, if we
explore the data in more detail, we can see that the fit of
$S_{iso1}+S_{iso2}$ model has an overall deviation to the data, and
single $S_d$ can roughly fit the data except in low-T region, where
the fitting result is higher than the experimental data. The
$S_{iso}+S_d$ model seems to give the best fitting in the whole
temperature region. The multigap scenario ($S_{iso}+S_d$) with a
very small s-wave gap of about 2 meV, which accounts for less than
30\% of the total quasiparticle DOS, seems possible. The global
fitting with $S_{iso}+S_d$, together with the relationship
$\gamma_e=\gamma_0+\alpha T^2$ in the low-T limit, suggest that
anisotropic gaps, possibly with nodes, may exist in these FeAs-based
superconductors.

\begin{figure}
\includegraphics[width=8cm]{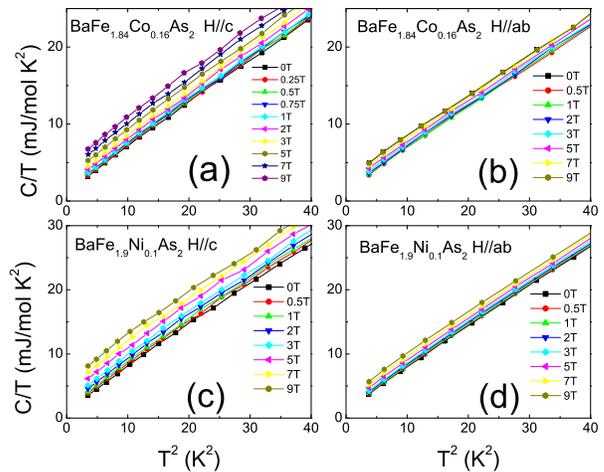}
\caption {(color online) Temperature dependence of the specific heat
C/T in the low temperature region for
BaFe$_{1.84}$Co$_{0.16}$As$_{2}$ and BaFe$_{1.9}$Ni$_{0.1}$As$_2$
with the magnetic field aligned with FeAs-plane and c-axis. }
\label{fig4}
\end{figure}

\begin{table}
\caption{Fitting parameters with different models for the Co-doped
sample. }
\begin{tabular}
{ccccccc}\hline \hline
model & $\Delta_1$(meV) & fraction-1   & $\Delta_2$(meV) & fraction-2\\
\hline
$S_{iso}$  & 4.2  & 100\%    &  -     &  -\\
$S_d$  & 5.4  & 100\%    &  -     &  -\\
$S_{iso}+S_d$       & 2.4  & 17\%     &  5.9   & 83\% \\
$S_{iso1}+S_{iso2}$       & 1.0  & 25\%     &  4.25  & 75\%\\
 \hline \hline
\end{tabular}
\caption{Fitting parameters with different models for the Ni-doped
sample. }
\begin{tabular}
{ccccccc}\hline \hline
model & $\Delta_1$(meV) & fraction-1   & $\Delta_2$(meV) & fraction-2\\
\hline
$S_{iso}$  & 3.1 & 100\%     &  -&  - \\
$S_d$ &4.1 & 100\%     &   -&  -\\
$S_{iso}+S_d$  & 2.0 & 29\%     &  4.7 & 71\% \\
$S_{iso1}+S_{iso2}$  & 1.15 & 29\%     &   3.3 & 71\%\\
 \hline \hline
\end{tabular}
\label{tab.1}
\end{table}

To explore whether the quadratic temperature dependence
$\gamma_e=\gamma_0+\alpha T^2$ is due to the presence of line nodes,
we measured also the magnetic field dependence of $\gamma_e$. The
data are plotted as $C/T$ vs. T$^2$ in Fig.4. One can see the
roughly linear behavior in the low-T region. It is clear that the
field induced enhancement of the electronic SH is larger when the
field is along c-axis than along FeAs-plane, indicating that the
upper critical field is higher along FeAs-plane than along
c-axis.\cite{WangZSPRB} By doing a linear extrapolation of the data
in Fig.4 to zero K, we get the field dependence of the electron SH
coefficient $\gamma_e(H)$, as shown in the upper part of Fig.5. It
is clear that $\gamma_e(H)$ increase linearly with magnetic field
for the two samples in both alignments. This excludes both the
vertical (like a d-wave) or horizontal line nodes (bottom-left
picture in Fig.5), since otherwise, as found in cuprate
superconductors, a square root relation $\gamma_e(H) \propto
\sqrt{H}$\cite{Volovik} in the clean limit, or a curved feature
$\gamma_e(H) \propto H/H_{c2}log(BH_{c2}/H)$ with impurity
scattering\cite{Hirschfield} should be observed.

\begin{figure}
\includegraphics[width=8cm]{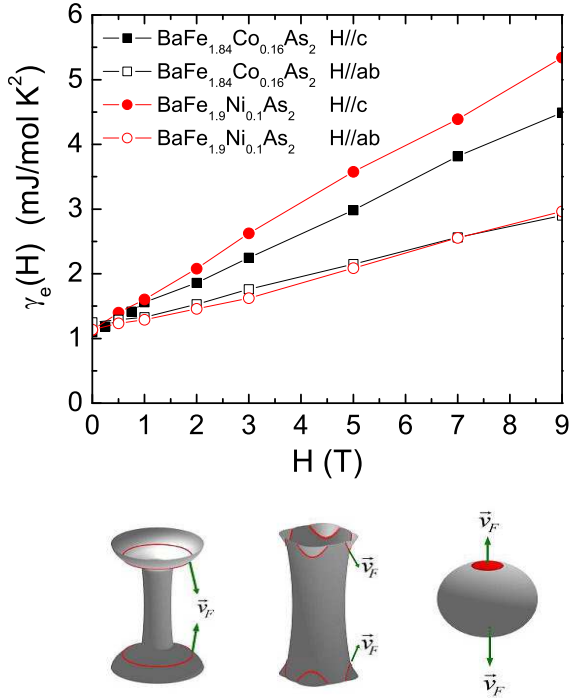}
\caption {(color online) Magnetic field dependence of the electronic
specific heat for the samples BaFe$_{1.84}$Co$_{0.16}$As$_{2}$ and
BaFe$_{1.9}$Ni$_{0.1}$As$_2$ when the field is aligned along
FeAs-plane and c-axis. A rough linear feature was observed in all
four cases. The bottom part illustrates the possible gap nodes on
the $\Gamma$ Fermi surfaces: (left) horizontal nodal lines; (middle)
small segments of nodal lines (shown by the red lines); (right)
point nodes on the 3D Fermi pockets.} \label{fig5}
\end{figure}

Although the quadratic temperature dependence
$\gamma_e=\gamma_0+\alpha T^2$ can also be attributed to the DOS
induced by the impurity scattering within the $S^\pm$
model\cite{Bang,Kontani,Gorden}, we would argue that this is
unlikely for the following reasons. Firstly, the exponent "2" here
may be achieved in the case of strong pair breaking\cite{Gorden},
which is in contrast to the small value of $\gamma_0$ $\sim$ 1.5
$mJ/mol K^2$ ($\leq 7.5\%$ $\gamma_n$). Furthermore, according to
Onari et al.,\cite{Kontani}, even with such a small amount
impurities (assuming to induce the large momentum scattering), the
superconductivity may not survive within the $S^\pm$ model.
Secondly, it is not straightforward to understand the linear field
dependence of $\gamma_e(H)-\gamma_0 \propto H$ with this picture. In
order to reconcile the observations of a small $\gamma_0$,
$\gamma_e=\gamma_0+\alpha T^2$ and $\gamma_e(H)-\gamma_0 \propto H$,
we propose that nodes, in the form of point-like or small segments,
may exist in our present samples. As indicated by the five-band
tight-binding calculations,\cite{Graser} small segments of nodal
lines (or called as accidental nodes) may exist on the hole-pockets
near $k_z = \pm \pi$, which is depicted by the middle-bottom cartoon
picture of Fig.5. An alternative interpretation would assume the
extreme case of point-like nodes with a closed 3D Fermi pocket. It
is natural that the pairing interaction by mediating the AF spin
fluctuations should be very weak on the right top or bottom of these
3D pockets where the Fermi velocity is along c-axis (see the
bottom-right picture of Fig.5), therefore the gap amplitude becomes
negligible. Actually these two pictures can reconcile all the
observations in this work. In a system with point nodes (or small
segments of line nodes), it is anticipated that $N_F\propto E^2$,
since the Doppler shift energy of $E_H\propto \sqrt{H}$, we have
$N_F\propto H$. Meanwhile, the relation $\gamma_e=\gamma_0+\alpha
T^2$ is also anticipated by the point-like nodes.

In summary, the electronic specific heat on the optimally Ni- and
Co-doped superconducting samples was derived by successfully
removing the phonon contributions. A quadratic temperature
dependence of $\gamma_e=\gamma_0 +\alpha T^2$ in the low-T limit was
discovered. The global temperature dependence of $\gamma_e$ can be
fitted with a two-component model with possible nodes. However the
linear field dependence $\gamma_e(H)\propto H$ observed with the
magnetic field along FeAs-plane and c-axis excludes the existence of
either vertical or horizontal line nodes. The results put strong
constraint on the pairing gap and can be reconciled by the model of
small segments of line nodes or point-like nodes.

\begin{acknowledgments}
We appreciate the discussions with P. Hirschfeld, S. Graser, A.
Chubukov, Y. Matsuda, G. Stewart, G. Q. Zheng and Z.
Tes$\check{a}$novi$\acute{c}$. This work is supported by the NSF of
China, the Ministry of Science and Technology of China (973
projects: 2006CB601000, 2006CB921107, 2006CB921802), and Chinese
Academy of Sciences within the knowledge innovation programm.
\end{acknowledgments}

$^{\star}$ hhwen@aphy.iphy.ac.cn


\begin{thebibliography}{99}

\bibitem{Kamihara2008} Y. Kamihara \emph{et al.}, J. Am. Chem. Soc. {\bf130}, 3296 (2008).
\bibitem{Mazin} I. I. Mazin \emph{et al.}, Phys. Rev. Lett. {\bf101}, 057003 (2008).
\bibitem{Kuroki} K. Kuroki \emph{et al.}, Phys. Rev. Lett. {\bf101}, 087004 (2008).
\bibitem{LeeDH} F. Wang \emph{et al.}, Phys. Rev. Lett. {\bf102}, 047005 (2009).
\bibitem{LiJX}Z. J. Yao, J. X. Li, Z. D. Wang, New J. Phys. {\bf11}, 025009 (2009).
\bibitem{LaFePOCarrington}J. D. Fletcher \emph{et al.}, Phys. Rev. Lett. {\bf102}, 147001 (2009).
\bibitem{LaFePO} C. W. Hicks \emph{et al.}, Phys. Rev. Lett. {\bf103}, 127003 (2009).
\bibitem{HDing} H. Ding \emph{et al.}, Europhys. Lett. {\bf83}, 47001 (2008).
\bibitem{Hashimoto}K. Hashimoto \emph{et al.}, Phys. Rev. Lett. {\bf102}, 017002 (2009).
\bibitem{Grafe}H.-J. Grafe \emph{et al.}, Phys. Rev. Lett. {\bf101}, 047003 (2009).
\bibitem{LuoXG}X. G. Luo \emph{et al.}, Phys. Rev. B {\bf 80}, 140503(R) (2009).
\bibitem{Prozorov}R. T. Gordon \emph{et al.}, Phys. Rev. Lett. {\bf102}, 127004
(2009). C. Martin \emph{et al.}, Phys. Rev. B {\bf80},
020501(R)(2009).
\bibitem{Matsuda} K. Hashimoto \emph{et al.}, arXiv:0907.4399 (2010). Y. Nakai \emph{et al.}, Phys. Rev. B {\bf 81}, 020503(R)(2010).
\bibitem{Sato} T. Sato \emph{et al.}, J. Phys. Soc. Jpn. {\bf77}, 063708 (2008).
\bibitem{MuG} G. Mu \emph{et al.}, Chin. Phys. Lett. {\bf25}, 2221 (2008). G. Mu \emph{et al.}, Phys. Rev. B {\bf79}, 174501 (2009).
\bibitem{Prozorov2}C. Martin \emph{et al.}, Phys. Rev. B {\bf 81}, 060505(R)(2010).
\bibitem{Taillefer}J. Ph. Reid  \emph{et al.} arXiv:1004.3804.
\bibitem{Graser} S. Graser, T. A. Maier, P. J. Hirschfeld, D. J. Scalapino, New J. Phys. {\bf11}, 025016 (2009).
\bibitem{FangLei} L. Fang \emph{et al.}, Phys. Rev. B \textbf{80}, 140508(R) (2009).
\bibitem{LuoHQ} M. Y. Wang \emph{et al.}, arxiv: 1002.3133 (2010).
\bibitem{Stewart} J. S. Kim \emph{et al.}, Phys. Rev. B81, 214507 (2010).
\bibitem{Keimer} P. Popovich \emph{et al.}, arXiv:1001.1074 (2010).
\bibitem{Ronning} K. Gofryk \emph{et al.}, Phys. Rev. B {\bf81},
184518(2010).
\bibitem{Imai}F. L. Ning \emph{et al.}, Phys. Rev.
Lett. \textbf{104}, 037001 (2010).
\bibitem{WangZSPRB}Z. S. Wang \emph{et al.}, Phys. Rev. B \textbf{78}, 140501(R) (2008).
\bibitem{Volovik} G. E. Volovik \emph{et al.}, JETP Lett. \textbf{58}, 469 (1993).
\bibitem{Hirschfield}  C. K\"{u}bert and P. J. Hirschfeld, Solid State
Commun. \textbf{105}, 459 (1998).
\bibitem{Bang}Y. Bang \emph{et al.}, Phys. Rev. B \textbf{79}, 054529 (2009).
\bibitem{Kontani}S. Onari and H. Kontani, Phys. Rev. Lett. \textbf{103},
177001 (2009)
\bibitem{Gorden}R. T. Gorden \emph{et al.}, Phys. Rev. B \textbf{81}, 180501(R) (2010).

\end{thebibliography}
\end{document}